\documentclass[manuscript]{aastex62}

\newcommand{\speed}[1]{#1 km~s${}^{-1}$}

\newcommand{\nfig}[1]{Figure~\ref{#1}}
\accepted{February 10, 2021}
\submitjournal{ApJ}

\shorttitle{MAGNETIC COUPLING ERUPTIONS IN THE SOLAR ATMOSPHERE}
\shortauthors{Chen et al.}

\begin{document}
\title{Direct Observation of A Large-scale CME Flux Rope Event Arising from an Unwinding Coronal Jet}

\correspondingauthor{Hechao Chen}
\email{hechao.chen@pku.edu.cn}
\author[0000-0001-7866-4358]{Hechao Chen}
\affil{School of Earth and Space Sciences, Peking University, 100871 Beijing, People’s Republic of China}
\affil{Yunnan Observatories,Chinese Academy of Sciences, 396 Yangfangwang, Guandu District, Kunming, 650216, China}

\author{Jiayan Yang}
\affil{Yunnan Observatories,Chinese Academy of Sciences, 396 Yangfangwang, Guandu District, Kunming, 650216, China}
\affil{Center for Astronomical Mega-Science, Chinese Academy of Sciences, 20A Datun Road, Chaoyang District, Beijing, 100012, China}

\author{Junchao Hong}
\affil{Yunnan Observatories,Chinese Academy of Sciences, 396 Yangfangwang, Guandu District, Kunming, 650216, China}
\affil{Center for Astronomical Mega-Science, Chinese Academy of Sciences, 20A Datun Road, Chaoyang District, Beijing, 100012, China}

\author{Haidong Li}
\affil{Yunnan Observatories,Chinese Academy of Sciences, 396 Yangfangwang, Guandu District, Kunming, 650216, China}
\affil{Center for Astronomical Mega-Science, Chinese Academy of Sciences, 20A Datun Road, Chaoyang District, Beijing, 100012, China}

\author{Yadan Duan}
\affil{Yunnan Observatories,Chinese Academy of Sciences, 396 Yangfangwang, Guandu District, Kunming, 650216, China}
\affil{Center for Astronomical Mega-Science, Chinese Academy of Sciences, 20A Datun Road, Chaoyang District, Beijing, 100012, China}
\affil{University of Chinese Academy of Sciences, 19A Yuquan Road, Shijingshan District, Beijing 100049, China}



\begin{abstract} 
Coronal mass ejections (CMEs) and coronal jets are two types of common solar eruptive phenomena, which often independently happen at different spatial scales. In this work, we present a stereoscopic observation of a large-scale CME flux rope arising from an unwinding blowout jet in a multipolar complex magnetic system. Based on a multi-band observational analysis, we find that this whole event starts with a small filament whose eruption occurs at a coronal geyser site after a series of homologous jets. Aided by magnetic field extrapolations, it reveals that the coronal geyser site forms above an elongate opposite-polarity interface, where the emergence-driven photospheric flux cancellation and repetitive reconnection are responsible for those preceding recurrent jets and also contribute to the ultimate filament destabilization. By interacting with overlying fields, the erupting filament breaks one of its legs and results in an unwinding blowout jet. Our estimation suggests that around 1.4$-$2.0 turns of twist release in its jet spire. This prominent twist transport in jet spire rapidly creates a newborn larger-scale flux rope from the jet base to a remote site. Soon after its formation, this large-scale flux rope erupts towards the outer coronae causing an Earth-directed CME. In its source region, two sets of distinct post-flare loops form in succession, indicating this eruption involves two-stage of flare magnetic reconnection. This work not only reveals a real magnetic coupling process between different eruptive activities but provides a new hint for understanding the creation of large-scale CME flux ropes during the solar eruption. 
\end{abstract}



\section{Introduction} \label{sec:intro}
Coronal mass ejections (CMEs) are the most violent eruptive magnetic activities in the solar atmosphere. They often appear as stellar-sized magnetized plasma bubbles in white-light coronagraph observations, and can rapidly release a vast amount of mass and energy into the inner heliosphere making our near-earth space a hazardous place \citep[e.g.,][]{2011LRSP....8....1C,2013AdSpR..51.1967S}. To date, almost all CME theories suggested that the core of CMEs correspond to rapidly erupting magnetic flux ropes (MFRs) \citep[e.g.,][]{2000JGR...10523153F,2000JGR...105.2375L}. As the desirable progenitor of CME flux ropes, an MFR is defined as a set of highly coherent helical magnetic field lines winding around one common axis. In such a non-potential topology, a mature MFR often appears with high storage of magnetic free energy and helicity, and is prone to suffer ideal MHD instabilities. Hence, the present of a mature MFR prior to or even during CME initiations is always given great attention. 
\par
In the past two decades, a lot of effort has been made in numerical simulations and observations on the formation and destabilization of active-region MFRs \citep[see reviews i.e.][]{2017ScChD..60.1383C,2019RSPTA.37780094G,2020SSRv..216..131P}. 
One important reason is that active-region MFR proxies often repeatedly form in newly-emerged bipolar active regions (ARs) and their eruptions can result in recurrent CMEs and flares in rapid succession \citep[e.g.,][]{2002ApJ...566L.117Z,2012A&A...539A...7L,2017ApJ...844..141L}.  
In accord with filament channel formations, it is widely believed that MFRs can be created along PILs in two ways: (1) the directly emergence of a mature MFR from the below photosphere \citep[e.g.,][]{2004ApJ...609.1123F,2004ApJ...610..588M,2008ApJ...673L.215O} or (2) successive reconnection of coronal fields induced by the continuously photospheric shearing or converging flows \citep[e.g.,][]{2000ApJ...529L..49A,2010ApJ...708..314A,2013ApJ...779..157G}. Due to differences in the spatial height of reconnection, the second one may respectively manifest as tether-cutting reconnection just at the onset of the eruption \citep[e.g.,][]{2010ApJ...725L..84L,2018ApJ...869...78C,2019ApJ...887..118C} or
slow flux cancellation in the photosphere long prior to the eruption \citep[e.g.,][]{2011A&A...526A...2G,2012ApJ...759..105S,2016ApJ...816...41Y}. In particular, apart from appearing in the formation phase of MFRs and filaments, flux cancelation is also thought as a popular triggering mechanism for CMEs and coronal jets \citep[e.g.,][]{2001ApJ...548L..99Z,2001ApJ...554..474Z,2011A&A...526A...2G,2018ApJ...869...78C,2018ApJ...864...68S}.
\par  
In a bipolar configuration, the ideal MHD models suggest that the sudden onset of MFR eruptions is directly triggered by the ideal torus instability \citep{2004A&A...413L..27T,2010ApJ...718..433O}, ideal kink instability \citep{1979SoPh...64..303H,2006PhRvL..96y5002K}, and also catastrophe mechanisms \citep{1993ApJ...417..368I,2000JGR...105.2375L}. Instead, the non-MHD models emphasize that MFR eruptions tend to take place if pre-flare reconnection below MFRs reducing their overlying magnetic constraints \citep[e.g.;][]{2000ApJ...545..524C,2001ApJ...552..833M,2013ApJ...771L..30J,2014ApJ...797L..15C,2018ApJ...869...78C}.
However, real source regions of many large-scale CMEs often cover more broad spatial ranges, connecting two inter-coupled ARs, or even a cluster of ARs \citep[e.g.,][]{2006SoPh..239..257W,2006A&A...445.1133Z,2007SoPh..244...75W,2015SCPMA..58.5682W,2007SoPh..241..329Z}. Accordingly, their CME progenitors are more likely associated with larger-scale pre-eruption structures, such as transequatorial magnetic loops, inter-connecting filaments among ARs, and extended bipolar PILs \citep[e.g.,][]{2006A&A...445.1133Z,2019ApJ...873...23Z,2007SoPh..244...75W,2013ApJ...764...91S}.
Under these multipolar complex magnetic systems, CME initiations and eruptive dynamics are no an alone behavior of a bipolar AR anymore, instead, it may tied to magnetic coupling and instability in the whole magnetic flux system \citep{1999ApJ...510..485A,2007SoPh..244...75W,2015SCPMA..58.5682W,2008AnGeo..26.3077V,2013ApJ...773...93S,2020ApJ...905..150Z}. 
\par
In addition, multipolar complex magnetic systems often breed frequent sympathetic eruptive events \citep{2003ApJ...588.1176M,2008ApJ...677..699J,2011ApJ...738..179J,2011ApJ...739L..63T,2011JGRA..116.4108S}. Under such a common dome of magnetic flux system, a preceding filament eruption can easily weaken the magnetic constraint overlying other filaments and thus trigger sympathetic CMEs \citep[e.g.,][]{2012ApJ...745....9Y,2015ApJ...803...68Y,2012ApJ...750...12S,2016ApJ...820..126J,2020A&A...640A.101H}. In particular, several observations reported that in helmet-streamer configurations, coronal jets can also drive streamer-puff \citep{2005ApJ...635L.189B,2007ApJ...661..543M,2016ApJ...822L..23P} or narrow CMEs \citep[e.g.,][]{2002ApJ...575..542W,2009SoPh..259...87N,2011ApJ...738L..20H,2020ApJ...901...94J}, supporting a physically link exist between these two types of eruptive phenomena. Recently, \citet{2016ApJ...822L..23P} found that a series of coronal jets occurring at the edge of the AR 12192 resulted in homogenous bubble-shaped CMEs. Based on the the release of magnetic twist from jet-base field into large-scale coronal loops and related coronal dimming during the jet eruptions, they infer that the jet-guiding coronal loops eventually blowed out as low-speed CME bubbles due to increasing twist. 
\par
In this paper, we study the initiation of an Earth-directed CME from a multipolar complex magnetic system, in which a large-scale CME flux rope event is found to arise from an unwinding coronal jet via magnetic coupling. With multi-wavelength and stereoscopic observations from Solar Dynamics Observatory \citep[SDO,][]{2012SoPh..275....3P} and \textit{Solar Terrestrial Relations Observatories} (STEREO), this work presents a direct observation for the creation of a large-scale MFR by rapid twist transform in unwinding jet spire, and sheds some light on the complex magnetic coupling in large-scale CME initiations. The layout of the remaining paper is as follows: Section 2 gives the data and methods; Section 3 presents the observational analysis and results; Section 4 presents the conclusion and discussion.
\par
\section{Data and Methods} \label{sec:Data}
Multi-wavelength EUV imaging data obtained from Atmospheric Imaging Assembly (AIA) \citep{2012SoPh..275...17L} on board SDO and H$\alpha$ center images from the Global Oscillation Network Group (GONG) are used to study CME source activities. Line-of-sight (LOS) magnetograms from Helioseismic and Magnetic Imager (HMI) \citep{2012SoPh..275..207S} are applied to study the photospheric magnetic evolution in source region. Meanwhile, observations from the \textit{Reuven Ramaty High Energy Solar Spectroscopic Imager} \citep[RHESSI;][]{2002SoPh..210....3L} and the Nancay radio heliograph \citep[NRH;][]{1997LNP...483..192K} are also used. The solar rotation of all these solar disk imaging data was removed through registering to a proper reference moment (09:30 UT).
In addition, the CME and associated phenomena are detected from the inner to the outer corona with the combination observations from the Large Angle and Spectrometric Coronagraph \citep[LASCO;][]{1995SoPh..162..357B} on board the \textit{Solar and Heliospheric Observatory} and Extreme Ultraviolet Imager \citep[EUVI;][]{2004SPIE.5171..111W} on board STEREO-A and B. 
\par
Based on the Active Region Patches \citep[SHARPs;][]{2014SoPh..289.3549B} vector field products of AR 11515, the photospheric vertical electric current ($J_{z}$) in source region is computed from the direct observed horizontal field according to the Ampere's law: $\vec{J}_{z}=\frac{1}{\mu_{0}}(\bigtriangledown \times \vec{B})_{z}$. To reveal the pre-eruption magnetic field configuration, we ``pre-processed"  CEA vector magnetograms to best suit the force-free condition, and then used them as the photospheric boundary to conduct a series of NLFFF magnetic field modelling with weighted optimization method \citep{2000ApJ...540.1150W,2004SoPh..219...87W}. 
\par
With these NLFFF and PF magnetic extrapolation modelings, coronal magnetic free energy ($E_f$) can be computed within a certain volume $V$ as follow: $E_{f}=\int_{V} \frac{B_{N}^2}{8\pi} dV - \int_{V} \frac{B_{P}^2}{8\pi} dV$,
where the subscripts $N$($P$) denotes NLFFF(PF) extrapolated magnetic field \citep{2012ApJ...748...77S}.
In addition, newly-injected magnetic energy from the below photosphere is also estimated from the energy (Poynting) flux that cross the photospheric surface as follow \citep{2012ApJ...761..105L}:
$\frac{d E_{in}}{dt}=\frac{1}{4\pi}\int_{S} B^{2}_{t}{V_{\bot n}} dS-\frac{1}{4\pi}\int_{S} (\textbf{\emph{B}}_{t}\cdot{\textbf{\emph{V}}_{\bot t}})B_{n} dS,$
where the emerge term (first term) comes from the emergence of magnetic tubes from the solar interior, and the shear term (second term) is generated by shear motions on the solar surface.
\section{Observational Results} \label{sec:Obs}
\begin{figure}[htb!]    
   \centerline{\includegraphics[width=0.8\textwidth,clip=]{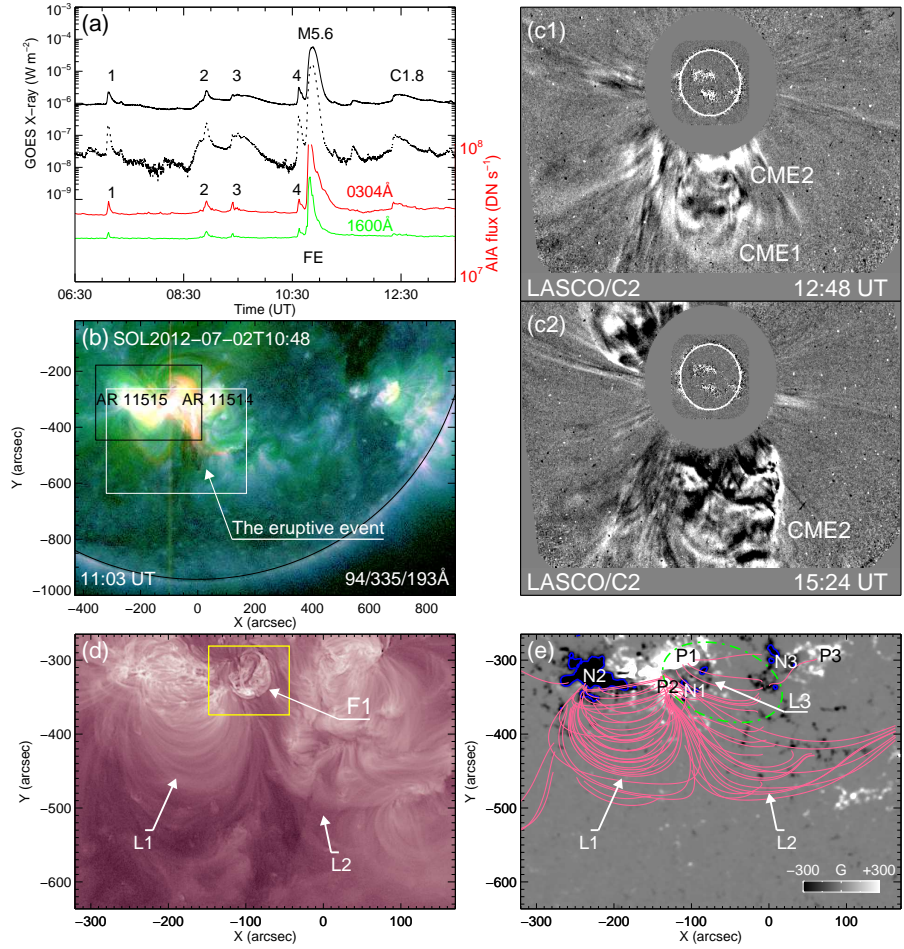}
              }
              \caption{(a) The temporal change of GOES flux in 1-8 \AA \ and 0.5-4 \AA , and the associated AIA flux that computed in panel (d). The numbers orderly mark four preceding emission enhancements before the major flare, which correspond to four recurrent jets before the main eruption. (b) The overview of the major solar eruption (SOL2012-07-02T10:48). (c) Two related CMEs, in which CME2 is closely related to our study.  (d) and (e): The overall magnetic environment of this event. Their FOVs are identical and denoted by the white box in panel (b). The related field lines, including ``L1", ``L2", and ``L3", are well traced by PFSS technique, and the former two can be observed from AIA 211 \AA \ image; the green elliptical dashed line denotes ``L3", corresponding to a latter mentioned fan-spine configuration. The yellow box in panel (d) denotes the field of view (FOV) of \nfig{fig3}. Other features of interest are marked in each panel; see the text for details.}
   \label{fig1}
\end{figure}
\par
The multipolar flux magnetic system that bred the solar eruptive event of our interest consists of two inter-coupled ARs (NOAA 11514 and 11515) (see \nfig{fig1}(b)).  As shown in \nfig{fig1}(a), the major flare (SOL2012-07-02T10:48) belongs to a short-duration one, which started, peaked, and rapidly ended at around 10:43, 10:48, and 10:57 UT, respectively. Previously, \citet{2014A&A...562A.110L} studied the sunspot splitting of AR 11515 and emphasized its role in triggering the major flare. Different from their study,  here we mainly investigate CME initiations and related magnetic coupling processes. Note that this solar eruption simultaneously resulted in two successive Earth-directed CMEs (see \nfig{fig1}(c1-c2)), but the second is the one of our interest since it is closely related to the formation and eruption of a large-scale CME flux rope.
\par
From 211 \AA \ observation and PF extrapolations in \nfig{fig1} (d) and (e), it can be seen that this multipolar flux system at least consisted of three groups of opposite-polarity magnetic fluxes: $P1$-$N1$, $P2$-$N2$, and $P3$-$N3$. As two bundles of high-lying coronal loops, $L1$ connected $N2$ and $P2$, while $L2$ connected $P1/P2$ and remote negative background magnetic fluxes; Instead, $L3$ connected $P1$ to $N3$ and disperse background fields. At the adjacent footpoints of $L1$ and $L2$, a small solar filament (F1) a length of $\sim$ 45 Mm resided in a newly emerged magnetic system ($P1$-$N1$). 
\par
\begin{figure}[htb!]    
   \centerline{\includegraphics[width=0.6\textwidth,clip=]{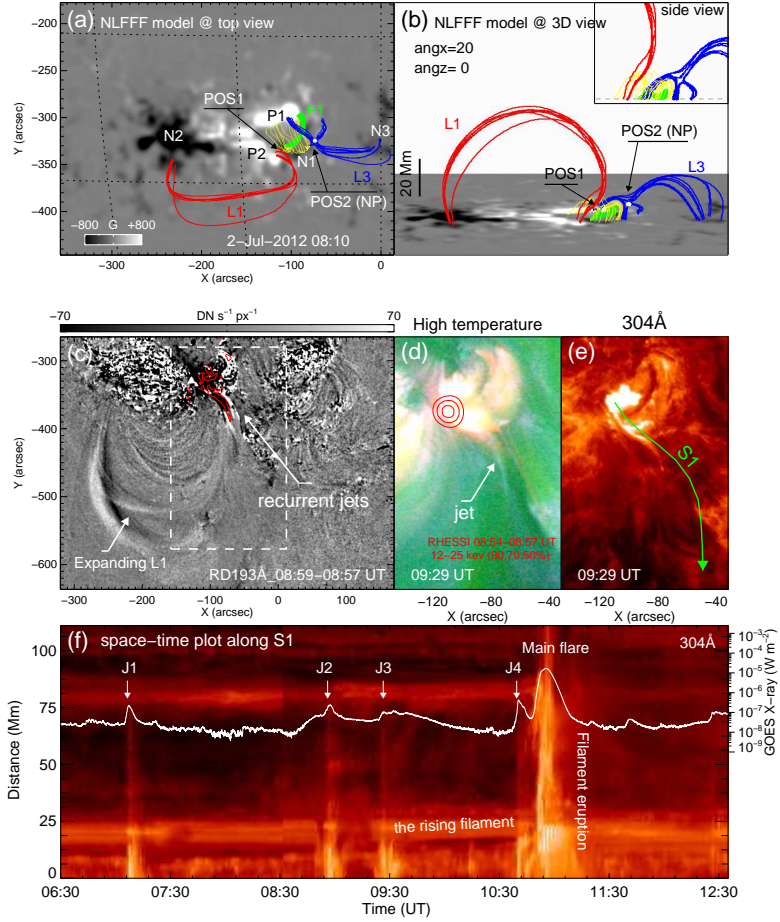}
              }
              \caption{(a) and (b): Selected field lines from the NLFFF extrapolation seen from a close-up 3D view, and its background is the B$_z$ map at Z = 0. (c): AIA 193 \AA \ running-difference image presents the jet-induced coronal loop expansion. The red contour traces the recurrent jets that occurred at the footpoint of $L1$. (d) The high temperature composite image of coronal jet, which blends 94, 193 and 335 \AA \ images. (e) AIA 304 \AA \ image observed during recurrent jets. The FOV of (d) and (e) are denoted by the white dash box in panel (c). (f) AIA 304 \AA \ space-time plot along the green S1 in panel (e), in which the GOES 1.0-8.0 \AA \ flux is also plotted for comparison. Features of interest are marked in each panel; see the text for details. An animation of panels (c), (d), (e), and (f) is available. The animation covers 06:30 UT to 12:30 UT with 60 s cadence. The video duration is 14 s.}
                 \label{fig2}
\end{figure}

Via NLFFF modeling, the pre-eruption magnetic configuration above the source region are better presented in \nfig{fig2}(a) and (b). In 3D perspectives, one can notice that there exist two privileged positions for the occurrence of magnetic reconnection: an opposite-polarity interface in POS1 and  a null point in POS2, according to 3D reconnection theory \citep{1994A&A...285.1023D}. POS1 corresponds to the interface between $L1$ and the emerging magnetic system $P1$-$N1$. Within this system, the filament structure of F1 is mainly constrained by a bundles of arch loops (yellow lines) that connects $P1$ and $N1$. Accordingly, the maximum twist number of F1 is derived as $-0.88$ turns from the NLFFF extrapolation by the equation: $T_{w} =  \frac{1}{4\pi} \int_{L} \alpha dl $ \citep{2006JPhA...39.8321B}. Meanwhile, near the north section of the filament structure, there exists a null-point-type configuration (blue lines, $L3$), which arched its outer spine to the remote negative-polarity flux $N3$ and partially restrained the filament structure with its inner spine lines.
\par
Before the occurrence of the major eruption, a so-called coronal geyser site \citep{2020ApJ...891..149P} formed the interface at the POS1. As shown in \nfig{fig2} (c-e) and its animation, four recurrent coronal jets successively took place there with similar narrow jet spires and evident plasma heating signals (especially the HXR emission and chromospheric flaring patch in \nfig{fig2}(d) and (e)). Such analogous dynamics features suggest that these recurrent jets are homologous ones \citep[e.g.;][]{2015ApJ...815...71C,2018ApJ...852...10L,2019ApJ...887..154L,2020ApJ...891..149P}. 
\par
\begin{figure}[htb!]    
   \centerline{\includegraphics[width=0.6\textwidth,clip=]{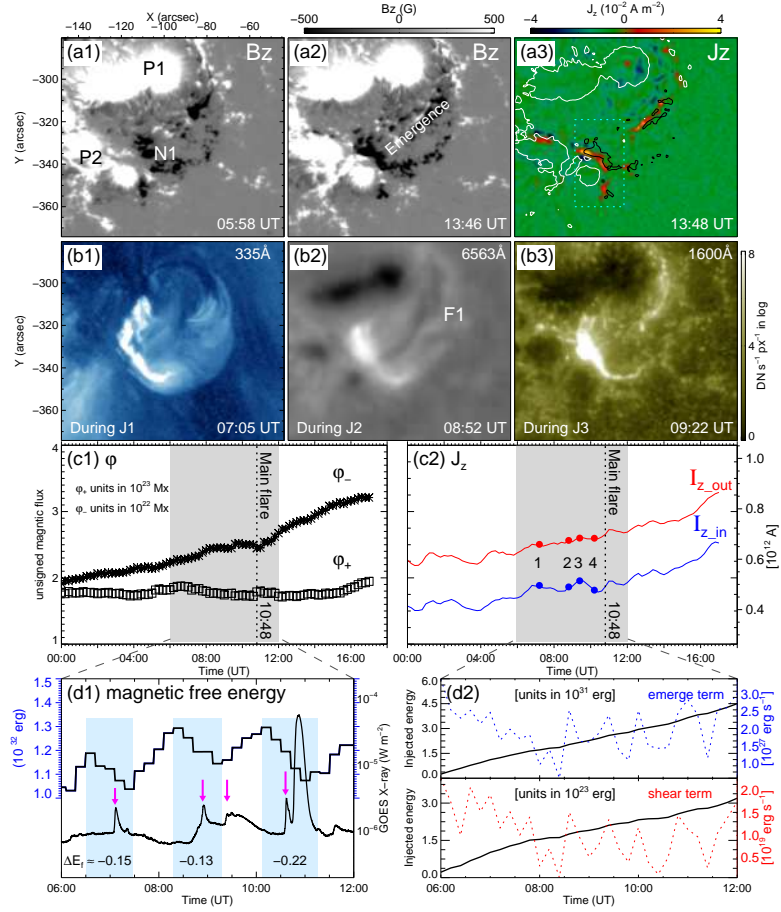}
              }
              \caption{Close-up view snapshots of the evolving source region (in the yellow dashed box in \nfig{fig1} (d)). (a1-a2): HMI $B_z$. (a3): Vertical current computed at the photosphere. (b1-b3) AIA 335\AA \ , GONG H$_{\alpha}$, and UV 1600 \AA \ observations. (c1): The unsigned magnetic fluxes computed in the source region. (c2): The profile of integral unsigned current $I_{z out}$ and $I_{z in}$ in the cyan dotted box of panel (a3). (d1): Computed magnetic free energy (\textit{E$_f$}) in the source region with time. (d2): Poynting flux cross the photospheric surface in the source region with time. Features of interest are marked in each panel; see the text for details.}
   \label{fig3}
\end{figure}

On the photosphere, as shown in \nfig{fig3}, the ongoing emerged $N1$ kelp converging toward $P2$ forming an emergence-driven canceling interface between $N1$ and $P2$, which is spatially coincide with POS1. During 00:00 to 18:00 UT,  $N1$ successively increased its flux from $1.9\times 10^{22}$ to $3.2\times 10^{22}$ Mx, while the positive flux demonstrated a weak decrease from 06:00 to 12:00 UT (see \nfig{fig3}(d1).
Moreover, along this canceling interface, enhanced vertical current flux density $J_z$ concurrently built up with an elongated pattern (see \nfig{fig3}(a3)). The integrated curves of its unsigned vertical current, $I_{out}$ and $I_{in}$, both demonstrate an increasing trend during this 18 hrs (see \nfig{fig3}(c2)). As responses of recurrent jets, episodes of analogous flaring patches took place along the localized $J_z$ enhancement. These results support that recurrent jets were triggered by repetitive reconnection at POS1 \citep[e.g.,][]{2010A&A...512L...2A,2013A&A...555A..19G}.  
\par
\par
With the occurrence of recurrent jets, the highly-lying $L1$ underwent sudden expansions at the same time (see \nfig{fig2}(a) and animation), implying a sudden reconstruction of magnetic topology. More importantly,  soon after the appearance of the 4th jet, F1 suddenly erupt from coronal geyser site at around 10:45 UT. To better describe this process, a space-time plot is made along the green slice 1 in \nfig{fig2}(e), which both goes through the southeast end of F1 and the ejecting trajectory of coronal jets. The result is presented in \nfig{fig2}(f) with the comparison of \textit{GOES} flux curve. From which, one can intuitively see that before its final eruption: (1) four preceding recurrent jets correspond to the four peaks in the GOES flux curves and source-region AIA fluxes (also see \nfig{fig1}(a));  (2) with the occurrence of recurrent jets, F1 displayed a quasi-static ascent. This indicates that such preceding flux peaks and recurrent jets might be considered as precursors for the imminent main eruption.
\par
As shown in in \nfig{fig3}(d1), during the occurrence of recurrent jets and the main eruption, the time profile of magnetic free energy in source region underwent three obvious drops. Thereinto, the former two respectively released $1.5 \times 10^{31}$ and $1.3 \times 10^{31}$ erg of free energy, while the last drop that includes the main eruption released  $2.2 \times 10^{31}$ erg of free energy. Meanwhile, mainly due to the emergence term of poynting flux, around $4.5 \times 10^{31}$ erg of magnetic energy was injected upward from the below photospheric surface and refilled such free energy decrease (see \nfig{fig3}(d2)). 
\par
\begin{figure}[htb!]    
   \centerline{\includegraphics[width=0.7\textwidth,clip=]{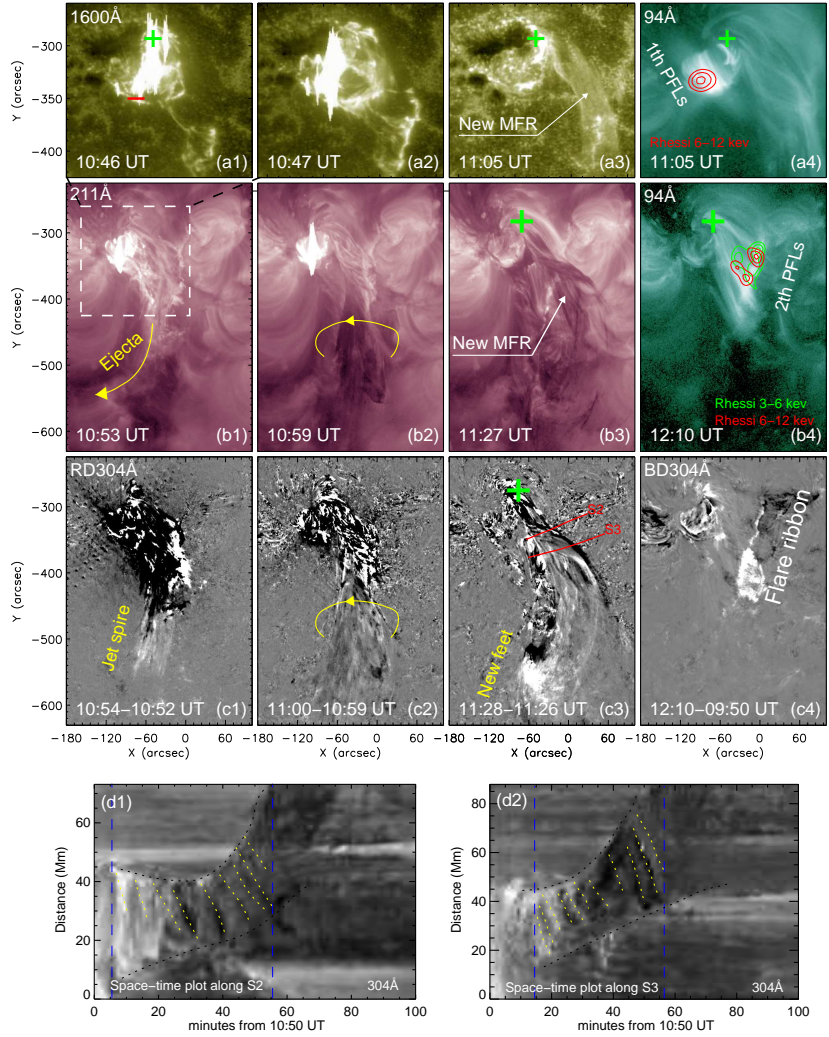}
              }
              \caption{(a1-a3): The ascent and disintegration of the filament observed in AIA 1600 \AA \ images. (b1-b3) and (c1-c3): AIA 211 \AA \ images and running-difference 304 \AA \ images. (a4) and (b4): Selected AIA 94 \AA \ images illustrate post-flare loops that observed after the filament eruption. The first(second) post-flare loops formed above(outside) the source region. (c4): A newborn flare ribbon observed in base-difference 304 \AA \ image. (d1) and (d2): space-time plots made along S2 and S3 in panel (c3). The white dash box in panel (b1) denotes the FOV of panels in the top row. Features of interest are marked in each panel; see the text for details. An animation of panels (b3),  (b4), and (c1) is available. The animation covers 10:30 UT to 12:29 UT with 60 s cadence. The video duration is 5 s.}
   \label{fig4}
\end{figure}
\par
Via interacting with its overlying NP configuration (POS2, see \nfig{fig2}(b)), the erupting F1 soon underwent a rapid disintegration and triggered the M5.6-class flare (see \nfig{fig4}(a1)-(a3)). As a result, typical signals of null-point reconnection, including a quasi-circular flare ribbon and a remote brightenings, were detected around the main flare ribbon \citep[e.g.;][]{2009ApJ...700..559M,2016ApJ...827...27Z,2018Ap&SS.363...26L,2019ApJ...886L..34L}. As the filament disintegration proceeded, the north (also positive-polarity) feet of F1 remained line-tied at its original position, but its south feet appeared as an ``open'' jet spire. Accordingly, the filament plasma rapidly ejected towards the southeast possibly along $L1$ and also displayed a conspicuous anti-clockwise unwinding motion (see \nfig{fig4}(b1)-(b3) and (c1)-(c3)). Compared with those preceding narrow and collimated jets, this unwinding jet can be characterized as a so-called blowout jet due to its broader jet spire \citep{2010ApJ...720..757M}.
\par
Assuming the main axis of the unwinding jet spire can be regarded as a circular cylinder and its fine threads rotates rigidly, a rough twist calculation of such unwinding jet spire can be given \citep[e.g.;][]{2011ApJ...735L..43S,2012RAA....12..573C,2013RAA....13..253H,2015ApJ...814L..13L,2018ApJ...852...10L,2019FrASS...6...44L}.
In \nfig{fig4} (d1) and (d2), two space-time plots are made perpendicular to its jet spire (along the red slices S2 and S2 in \nfig{fig4}(c3)). one can see that rolling motions nearly perpendicular to the jet spire display as dark/bright strips, which last for 42 and 50 mins in (d1) and (d2), respectively.  By tracing and conducting linear fittings along these inclined dark/bright strips, the average rotational speed ($\nu_r$) of this unwinding jet can be estimated as \speed{48.5} and \speed{38.2}, respectively. The average width of the jet spire (23.09 Mm and 21.40 Mm) is equal to its diameter ($d$). Accordingly, their angular speeds ($\omega =2\nu_r /d$) along S2 and S3 are computed as $3.57\times10^{-3}$ rad s$^{-1}$ (period 1758 s) and $4.20\times10^{-3}$ rad s$^{-1}$ (period 1494 s), respectively. So, the total amount of twist released in this jet spire is roughly at 1.43$-$2.01 turns, or 2.86$-$4.02$\pi$. This result reaches an agreement with other previous twist estimation (1.2$-$4.7 turns) in comparable coronal jets \citep[e.g.,][]{2011ApJ...735L..43S,2012RAA....12..573C,2013ApJ...769..134M,2015ApJ...806...11M,2015ApJ...814L..13L,2019FrASS...6...44L}. Note that the median twist number of pre-eruption F1 is derived as 0.88 turns from the NLFFF extrapolation. A similar disagreement is also found in \citep{2018ApJ...852...10L}. We think the twist estimation by the feature tracing method is more reliable, because: (1) the feature tracing method involves less assumptions; (2) the low-lying F1 might require a sophisticated non-force-free modeling due to its small scale. 
\par
\begin{figure}[htb!]    
   \centerline{\includegraphics[width=0.8\textwidth,clip=]{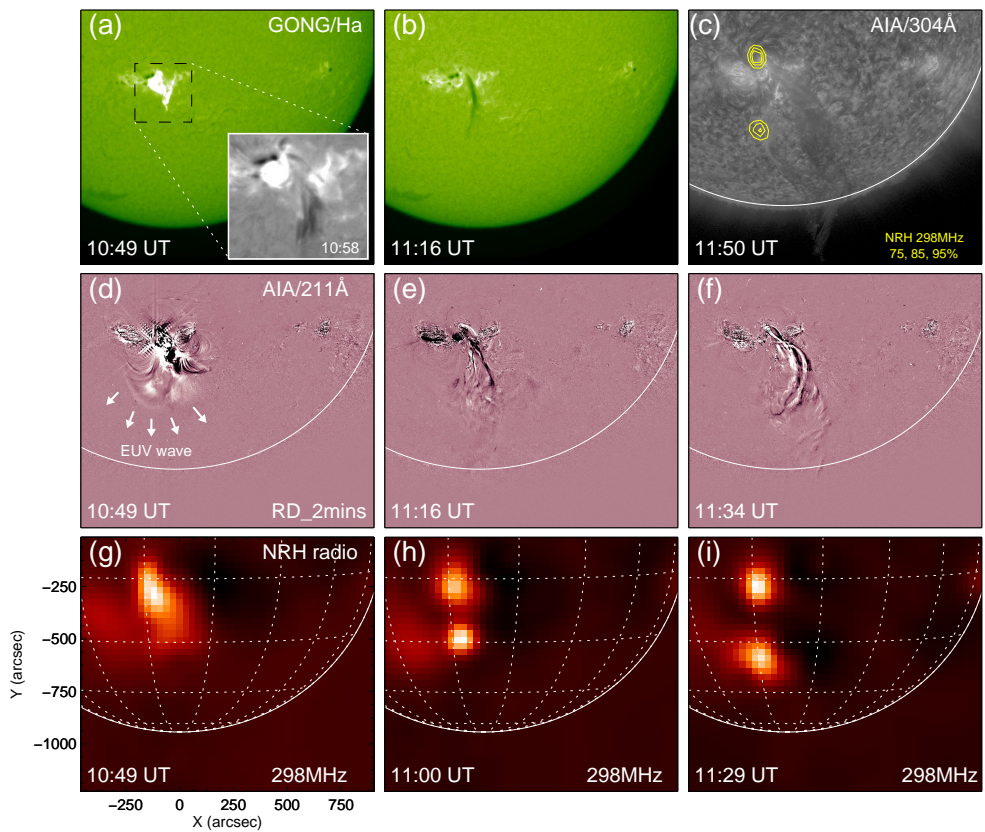}
              }
              \caption{Snapshots of the global eruption observed in (a-c): H$_{\alpha}$ and AIA 304\AA \ images, (d-f): running difference AIA 211 \AA \ images, and (g-i): NRH radio imaging observation at 298 MHz. Features of interest are marked in each panel; see the text for details.}
   \label{fig5}
\end{figure}
\par
\begin{figure}[htb!]    
   \centerline{\includegraphics[width=0.8\textwidth,clip=]{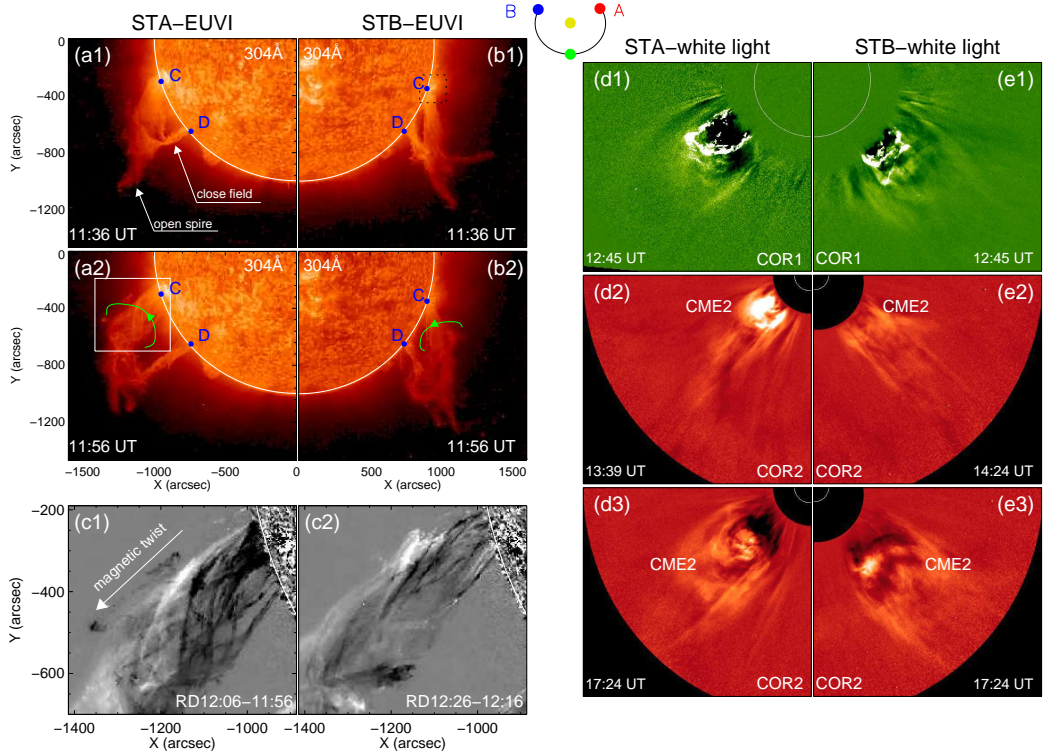}
              }
              \caption{STEREO observations on the newborn CME flux rope. The left part: The erupting filament in running-difference EUVI 304 \AA \ image. Close-up view snapshots of fine twisted structures at the north feet of the CME flux rope are presented in (c1-c2). The right part: The corresponding CME captured by white coronagraph COR1 and  COR2. Features of interest are marked in each panel; see the text for details.}
   \label{fig6}
\end{figure}

Due to the unwinding dynamics, a new filament feet gradually appeared within the ``open" jet spire and demonstrated an apparent slipping motion towards the south (see \nfig{fig4}(c3)). By the time of 11:28 UT, the jet-like plasma ejecta traced out a newborn close magnetic structure. With the inspection of AIA (see animation of \nfig{fig4}) and STEREO-A/EUVI observations, we found that the newborn close magnetic structure actually corresponds to a newborn large-scale MFR. Especially near its north feet, prominent magnetic twist fields and anti-clockwise unwinding (rolling) motion indeed exist (see \nfig{fig6}(c1) and its animation). It must be pointed out that the creation of this newborn large-scale MFR accomplished at a relatively high coronal height, thus it soon erupt upwards.
\par
\nfig{fig5} further presents this eruption process with a larger FOV. In H$_\alpha$ observations, this erupting large-scale MFR manifested as an obvious erupting filament against the solar disk, and rapidly disappeared by the time of 11:25 UT (see \nfig{fig5}(a) and (b)). In running-difference 211 \AA \ images, it is found that as the major flare triggered, an EUV wave propagated toward the south, which might be the coronal imprint of the first CME (CME1) (also see \nfig{fig1}(c1)). After the major flare, the large-scale MFR arose from the coronal geysers and slowly erupt toward southwest, which eventually led to the second CME (CME2). This jet-CME coupling eruption needs to differentiate from the simple extension of coronal jets in white-light observations \citep{2002ApJ...575..542W,2011ApJ...738L..20H,2020ApJ...901...94J} and other so-called twin CME events \citep{2012ApJ...745..164S,2019ApJ...881..132D,2019ApJ...877...61M}, because a newborn large-scale MFR was indeed created during the solar eruption by the rapid twist transport from jet base to  background fields.
\par
In particular, two unique features are found in the eruption of the large-scale MFR.
First, as mentioned before, the newborn south feet of the large-scale MFR showed an apparent ``drift" toward the south in 304 and 211 \AA \ imaging observations. This evolution behavior is also evidenced by simultaneous radio imaging observations at 298 MHz from NRH, which corresponds to a computed height of around 170 Mm above the photosphere based on coronal density model of \citet{1999ApJ...523..812S} and the plasma-density relationship ($f = 8.98 \times \sqrt{n_e} $). 
As energetic electrons are injected into and filled its erupting volume, the feet of the newborn large-scale MFR were clearly imaged as a pair of distinct radio sources in \nfig{fig5}(c). Similar feet signatures of eruptive MFRs were also reported by other previous radio imaging observations \citep[e.g.,][]{2020FrASS...7...79C,2020ApJ...895L..50C}, but the feet radio source usually remain stationary.
In this current observation, the north one remained stationary at its spatial position, while the south one first appeared at around 10:55 UT, and demonstrated a south-orientated ``drift" during 11:00 to 11:29 UT (see \nfig{fig5}(g-i)).
\par
Second, signatures of an extra flare reconnection was detected below the erupting large-scale MFR. As shown in \nfig{fig4}(b4) and (c4), by the time of 12:10 UT, a set of new post-flare loops (2th PFLs) formed behind the erupting large-scale MFR, bridging a newly-formed chromospheric ribbon and the main flare region. Accordingly, $RHESSI$ sources was also detected at the loop-top of new post-flare loops (see \nfig{fig5}(b4)), indicating the occurrence of particle acceleration. On the whole, the creation of large-scale MFR and the appearance of the 2th PFLs provide solid evidence for this jet-CME coupling eruption. In the GOES flux, as shown in \nfig{fig1}(a), the associated X-ray emission enhancement in 2th PFLs is also clearly detected as another an C-class flare. Together with the jet-driven M5.6-class main flare,  this result supports that this coupling erption event involves two-stage of flare magnetic reconnection.
\par
\par
With the aid of stereoscopy observations from STEREO, it becomes more evident in \nfig{fig6} that the blowout jet first appeared along the high-lying loop $L1$, and then the newborn large-scale MFR displayed as a giant prominence eruption with two feet rooted at the source region \textit{``C"} (marked by black dashed box) and a remote region \textit{``D"}, respectively. The distance between  \textit{``C"} and  \textit{``D"} spans more than 270 Mm. 
With a close-up inspection, one can even notice that an anti-clockwise rolling motion appeared in the north leg of the erupting large-scale MFR, which indicating an obvious transfer of magnetic twist took place from the feet \textit{``C"} to the another feet \textit{``D"} (also see \nfig{fig6}(c1-c2)). 
In the right part of \nfig{fig6}, the white light coronagraphs from STEREO cor1 and cor2 well recorded its related CME. CME2 first appeared at 13:22 UT and successfully propagated to more than 18 $R_s$ with a relative low velocity of around \speed{300}. 
\section{Conclusion and Discussion}
CMEs and coronal jets are two types of common solar eruptive phenomena, which often independently happen at different spatial scales. The former are well-known for stellar-sized violent magnetized plasma explosions and their potential capacity in causing hazardous space weather \citep[e.g.,][]{2011LRSP....8....1C,2013AdSpR..51.1967S}, while the latter, as ubiquitous smaller-scale eruptive phenomena, are best known for their important contributions to the heating and mass supply for the upper solar atmosphere \citep[e.g.,][]{2014Sci...346A.315T,2016SSRv..201....1R,2019Sci...366..890S}. In this paper, we study the origin of a large-scale CME flux rope event arising from an unwinding coronal jet. Based on the stereoscopic and multi-bands observational analysis, we find that this whole eruptive event started with a small-scale filament within a multipolar complex magnetic system whose eruption first triggered an unwinding blowout jet and an M5.6 short-duration flare. Due to the subsequent release of magnetic twist from the jet base, a newborn larger-scale MFR was then created in the unwinding jet spire, with its new south feet exchanged to a remote site (around 270 Mm far from the jet base). Finally, this newborn large-scale MFR successfully erupt into the outer coronae driving a stellar-sized Earth-directed CME, leaving another an C1.8 flare in its source region. On the whole, this event highlights the pathway of a real magnetic coupling process in the initiation of the Earth-directed CME, supporting the view that some large-scale coronal eruptive phenomena can originate from the magnetic coupling of different magnetic activities at various spatial scales \citep{2007SoPh..244...75W,2015SCPMA..58.5682W,2013ApJ...773...93S}.
\par
Despite that the release of magnetic twist phenomena are frequently detected in coronal jets \citep{2011A&A...532L...9C,2011ApJ...735L..43S,2012RAA....12..573C,2013RAA....13..253H,2013ApJ...769..134M,2015ApJ...806...11M,2015ApJ...814L..13L,2019ApJ...887..239Y,2019FrASS...6...44L} and filament eruption events \citep[e.g.;][]{2014ApJ...797...52Y,2020ApJ...904...15Y,2018MNRAS.476.1286J,2018ApJ...866...96J}, but their resultant consequence are poorly studied yet. This work provides a direct stereoscopic imaging observation on the creation of a large-scale erupting CME flux rope in an unwinding coronal jet. Different from previous mentioned flux-emergence formation mechanism \citep[e.g.,][]{2004ApJ...609.1123F,2004ApJ...610..588M,2008ApJ...673L.215O,2018SoPh..293...93C,2019ApJ...874...96Y} and reconnection-cancellation formation mechanism \citep[e.g.,][]{2012ApJ...759..105S,2017ApJ...839..128W,2018ApJ...859..148W,2018ApJ...869...78C,2019ApJ...887..118C}, this observation supports an alternative scenario: the rapid magnetic twist transport from the jet base to background fields directly result in the formation of newborn large-scale MFR. Moreover, a rough twist estimation of the blowout jet indicates that around 1.43$-$2.01 turns of magnetic twist was released in its jet spire. The twist characteristics of the newborn large-scale flux rope and its anti-clockwise rolling motion can also be clearly recognized in STEREO/EUVI observations. In addition, the location of its ``drift" remote feet is also confirmed as a moving radio source in simultaneous NRH radio observations. 
\par
Apart from our observation, applying a state-of-the-art of data-constrained 3D MHD simulations and observations, \citet{2018ApJ...866...96J} also noticed that in NOAA 11283, an erupting sigmoidal flux rope breaks one of its legs, and quickly gives birth to a new tornado-like flux rope that is highly twisted and has multiple connections to the Sun (see also a similar 3D simulation from \citet{2020ApJ...903..129P}, which focus on the same event but mainly studies its related coronal dimmings). \citet{2016ApJ...822L..23P} also reported a series of coronal jets occurring at the edge of the AR 12192, which resulted in homogenous bubble-shaped CMEs. Based on the release of magnetic twist from the jet-base field into large-scale coronal loops and related coronal dimming during the jet eruptions, \citet{2016ApJ...822L..23P} inferred that the jet-guiding coronal loops eventually blew out as low-speed CME bubbles due to increasing twist. Focusing on one of these CME-producing jets, \citet{2015ApJ...814L..13L} also provided a detailed twist estimation in the jet spire with high-resolution chromospheric observations, and they reported that the total rotation angle is high up to 8$\pi$. In the current work, the estimated rotation angle in the unwinding jet ranges from around 2.8$-$4.0$\pi$, which should be enough for the creation of the newborn CME flux rope. 
\par
Before the occurrence of the blowout jet, repetitive reconnection triggered four obvious recurrent jets above an elongate enhanced $J_z$ enhancement between opposite-polarity converging flux ($P2$ and $N1$) (namely at POS1 ). These recurrent jets demonstrate narrow jet spire and short lifetime, thus may termed as ``standard" jets, while the blowout jet of our interest has a broader unwinding spire and triggered by a microfilament eruption scenario \citep{2010ApJ...720..757M,2015Natur.523..437S}. In their source region, flux emergence provided a continual injection of poynting flux and magnetic free energy for recurrent eruptions (see \nfig{fig3}(d1-d2)). On the other hand, flux emergence also introduced an emergence-driven canceling interface. As a result, reconnection between F1-contained newly emerged flux and pre-existing field happened, which weaken its constraint above the F1 at some degree \citep{2007ApJ...669.1359S,2019ApJ...874...96Y}. Consistent with the result of \citet{2014A&A...562A.110L} and \citet{2016ApJ...821..100S}, these suggest that both flux emergence and its driven flux cancellation should play a role in the onset of this whole magnetic coupling eruption. 
\par
At last remark, two issues worthy of mention here. First, in order to fully understand such jet-CME coupling eruption events, the energetics and interplanetary effects of jet-driven CMEs should be linked back to their jet base parameters in the future, especially the twisting effect of pre-jet filaments. Second, it must be pointed out that this newborn large-scale CME flux rope in the present observation immediately erupt due to its higher formation height, despite that it actually created along an extended PIL\footnote{See the online synoptic magnetogram: \url{http://jsoc.stanford.edu/data/hmi/synoptic/hmi.Synoptic_Ml.2125.png}.}. Therefore, whether this twist-transport formation scenario can apply to explain the appearance of those long-term exited large-scale MFRs among inter-coupled ARs \citep[e.g.,][]{2019ApJ...873...23Z} remains unclear.


\acknowledgments
The authors sincerely thank the referee for constructive suggestions and comments. H.C.C. thanks Dr. Guiping Zhou for helpful discussions after the on-line seminar, ``Frontiers in solar physics", held by the Key Laboratory of Solar Activity of NAO. H.C.C. is supported by the National Postdoctoral Program for Innovative Talents (BX20200013) and China Postdoctoral Science Foundation (2020M680201); This work is also supported by the National Key R$\&$D Program of China (2019YFA0405000), and the National Natural Science Foundation of China under grants 11633008, 11873088, 11933009, and 11703084.
%

\vspace{5mm}


\begin{thebibliography}{}
\bibitem[Amari et al.(2000)]{2000ApJ...529L..49A} Amari, T., Luciani, J.~F., Mikic, Z., et al.\ 2000, \apjl, 529, L49. doi:10.1086/312444
\bibitem[Antiochos et al.(1999)]{1999ApJ...510..485A} Antiochos, S.~K., DeVore, C.~R., \& Klimchuk, J.~A.\ 1999, \apj, 510, 485. doi:10.1086/306563
\bibitem[Archontis et al.(2010)]{2010A&A...512L...2A} Archontis, V., Tsinganos, K., \& Gontikakis, C.\ 2010, \aap, 512, L2. doi:10.1051/0004-6361/200913752
\bibitem[Aulanier et al.(2010)]{2010ApJ...708..314A} Aulanier, G., T{\"o}r{\"o}k, T., D{\'e}moulin, P., et al.\ 2010, \apj, 708, 314. doi:10.1088/0004-637X/708/1/314
\bibitem[Bemporad et al.(2005)]{2005ApJ...635L.189B} Bemporad, A., Sterling, A.~C., Moore, R.~L., et al.\ 2005, \apjl, 635, L189. doi:10.1086/499625
\bibitem[Berger \& Prior(2006)]{2006JPhA...39.8321B} Berger, M.~A. \& Prior, C.\ 2006, Journal of Physics A Mathematical General, 39, 8321. doi:10.1088/0305-4470/39/26/005
\bibitem[Bobra et al.(2014)]{2014SoPh..289.3549B} Bobra, M.~G., Sun, X., Hoeksema, J.~T., et al.\ 2014, \solphys, 289, 3549. doi:10.1007/s11207-014-0529-3
\bibitem[Brueckner et al.(1995)]{1995SoPh..162..357B} Brueckner, G.~E., Howard, R.~A., Koomen, M.~J., et al.\ 1995, \solphys, 162, 357. doi:10.1007/BF00733434
\bibitem[Carley et al.(2020)]{2020FrASS...7...79C} Carley, E.~P., Vilmer, N., \& Vourlidas, A.\ 2020, Frontiers in Astronomy and Space Sciences, 7, 79. doi:10.3389/fspas.2020.551558
\bibitem[Curdt \& Tian(2011)]{2011A&A...532L...9C} Curdt, W. \& Tian, H.\ 2011, \aap, 532, L9. doi:10.1051/0004-6361/201117116
\bibitem[Chen et al.(2020)]{2020ApJ...895L..50C} Chen, B., Yu, S., Reeves, K.~K., et al.\ 2020, \apjl, 895, L50. doi:10.3847/2041-8213/ab901a
\bibitem[Chen et al.(2018)]{2018ApJ...869...78C} Chen, H., Duan, Y., Yang, J., et al.\ 2018, \apj, 869, 78. doi:10.3847/1538-4357/aaead1
\bibitem[Chen et al.(2018)]{2018SoPh..293...93C} Chen, H., Yang, J., Yang, B., et al.\ 2018, \solphys, 293, 93. doi:10.1007/s11207-018-1311-8
\bibitem[Chen et al.(2019)]{2019ApJ...887..118C} Chen, H., Yang, J., Ji, K., et al.\ 2019, \apj, 887, 118. doi:10.3847/1538-4357/ab527e
\bibitem[Chen et al.(2012)]{2012RAA....12..573C} Chen, H.-D., Zhang, J., \& Ma, S.-L.\ 2012, Research in Astronomy and Astrophysics, 12, 573. doi:10.1088/1674-4527/12/5/009
\bibitem[Chen et al.(2014)]{2014ApJ...797L..15C} Chen, H., Zhang, J., Cheng, X., et al.\ 2014, \apjl, 797, L15. doi:10.1088/2041-8205/797/2/L15
\bibitem[Chen et al.(2015)]{2015ApJ...815...71C} Chen, J., Su, J., Yin, Z., et al.\ 2015, \apj, 815, 71. doi:10.1088/0004-637X/815/1/71
\bibitem[Chen \& Shibata(2000)]{2000ApJ...545..524C} Chen, P.~F. \& Shibata, K.\ 2000, \apj, 545, 524. doi:10.1086/317803
\bibitem[Chen(2011)]{2011LRSP....8....1C} Chen, P.~F.\ 2011, Living Reviews in Solar Physics, 8, 1. doi:10.12942/lrsp-2011-1
\bibitem[Cheng et al.(2017)]{2017ScChD..60.1383C} Cheng, X., Guo, Y., \& Ding, M.\ 2017, Science China Earth Sciences, 60, 1383. doi:10.1007/s11430-017-9074-6
\bibitem[Demoulin et al.(1994)]{1994A&A...285.1023D} Demoulin, P., Henoux, J.~C., \& Mandrini, C.~H.\ 1994, \aap, 285, 1023
\bibitem[Duan et al.(2019)]{2019ApJ...881..132D} Duan, Y., Shen, Y., Chen, H., et al.\ 2019, \apj, 881, 132. doi:10.3847/1538-4357/ab32e9
\bibitem[Fan \& Gibson(2004)]{2004ApJ...609.1123F} Fan, Y. \& Gibson, S.~E.\ 2004, \apj, 609, 1123. doi:10.1086/421238
\bibitem[Forbes(2000)]{2000JGR...10523153F} Forbes, T.~G.\ 2000, \jgr, 105, 23153. doi:10.1029/2000JA000005
\bibitem[Georgoulis et al.(2019)]{2019RSPTA.37780094G} Georgoulis, M.~K., Nindos, A., \& Zhang, H.\ 2019, Philosophical Transactions of the Royal Society of London Series A, 377, 20180094. doi:10.1098/rsta.2018.0094
\bibitem[Green et al.(2011)]{2011A&A...526A...2G} Green, L.~M., Kliem, B., \& Wallace, A.~J.\ 2011, \aap, 526, A2. doi:10.1051/0004-6361/201015146
\bibitem[Guo et al.(2013)]{2013A&A...555A..19G} Guo, Y., D{\'e}moulin, P., Schmieder, B., et al.\ 2013, \aap, 555, A19. doi:10.1051/0004-6361/201321229
\bibitem[Guo et al.(2013)]{2013ApJ...779..157G} Guo, Y., Ding, M.~D., Cheng, X., et al.\ 2013, \apj, 779, 157. doi:10.1088/0004-637X/779/2/157
\bibitem[Hong et al.(2013)]{2013RAA....13..253H} Hong, J.-C., Jiang, Y.-C., Yang, J.-Y., et al.\ 2013, Research in Astronomy and Astrophysics, 13, 253. doi:10.1088/1674-4527/13/3/001
\bibitem[Hong et al.(2011)]{2011ApJ...738L..20H} Hong, J., Jiang, Y., Zheng, R., et al.\ 2011, \apjl, 738, L20. doi:10.1088/2041-8205/738/2/L20
\bibitem[Hood \& Priest(1979)]{1979SoPh...64..303H} Hood, A.~W. \& Priest, E.~R.\ 1979, \solphys, 64, 303. doi:10.1007/BF00151441
\bibitem[Hou et al.(2020)]{2020A&A...640A.101H} Hou, Y.~J., Li, T., Song, Z.~P., et al.\ 2020, \aap, 640, A101. doi:10.1051/0004-6361/202038348
\bibitem[Isenberg et al.(1993)]{1993ApJ...417..368I} Isenberg, P.~A., Forbes, T.~G., \& Demoulin, P.\ 1993, \apj, 417, 368. doi:10.1086/173319
\bibitem[Jiang et al.(2013)]{2013ApJ...771L..30J} Jiang, C., Feng, X., Wu, S.~T., et al.\ 2013, \apjl, 771, L30. doi:10.1088/2041-8205/771/2/L30
\bibitem[Jiang et al.(2018)]{2018ApJ...866...96J} Jiang, C., Feng, X., \& Hu, Q.\ 2018, \apj, 866, 96. doi:10.3847/1538-4357/aadd08
\bibitem[Jiang et al.(2011)]{2011ApJ...738..179J} Jiang, Y., Yang, J., Hong, J., et al.\ 2011, \apj, 738, 179. doi:10.1088/0004-637X/738/2/179
\bibitem[Jiang et al.(2008)]{2008ApJ...677..699J} Jiang, Y., Shen, Y., Yi, B., et al.\ 2008, \apj, 677, 699. doi:10.1086/529417
\bibitem[Joshi et al.(2016)]{2016ApJ...820..126J} Joshi, N.~C., Schmieder, B., Magara, T., et al.\ 2016, \apj, 820, 126. doi:10.3847/0004-637X/820/2/126
\bibitem[Joshi et al.(2018)]{2018MNRAS.476.1286J} Joshi, N.~C., Nishizuka, N., Filippov, B., et al.\ 2018, \mnras, 476, 1286. doi:10.1093/mnras/sty322
\bibitem[Joshi et al.(2020)]{2020ApJ...901...94J} Joshi, R., Wang, Y., Chandra, R., et al.\ 2020, \apj, 901, 94. doi:10.3847/1538-4357/abaf5a
\bibitem[Kerdraon \& Delouis(1997)]{1997LNP...483..192K} Kerdraon, A. \& Delouis, J.-M.\ 1997, Coronal Physics from Radio and Space Observations, 192. doi:10.1007/BFb0106458
\bibitem[Kliem \& T{\"o}r{\"o}k(2006)]{2006PhRvL..96y5002K} Kliem, B. \& T{\"o}r{\"o}k, T.\ 2006, \prl, 96, 255002. doi:10.1103/PhysRevLett.96.255002
\bibitem[Lemen et al.(2012)]{2012SoPh..275...17L} Lemen, J.~R., Title, A.~M., Akin, D.~J., et al.\ 2012, \solphys, 275, 17. doi:10.1007/s11207-011-9776-8
\bibitem[Li et al.(2012)]{2012A&A...539A...7L} Li, L.~P., Zhang, J., Li, T., et al.\ 2012, \aap, 539, A7. doi:10.1051/0004-6361/201015796
\bibitem[Li et al.(2019)]{2019ApJ...886L..34L} Li, H., Yang, J., Hong, J., et al.\ 2019, \apjl, 886, L34. doi:10.3847/2041-8213/ab564e
\bibitem[Li et al.(2018)]{2018Ap&SS.363...26L} Li, H., Yang, J., Jiang, Y., et al.\ 2018, \apss, 363, 26. doi:10.1007/s10509-017-3244-3
\bibitem[Li et al.(2015)]{2015ApJ...814L..13L} Li, X., Yang, S., Chen, H., et al.\ 2015, \apjl, 814, L13. doi:10.1088/2041-8205/814/1/L13
\bibitem[Lin \& Forbes(2000)]{2000JGR...105.2375L} Lin, J. \& Forbes, T.~G.\ 2000, \jgr, 105, 2375. doi:10.1029/1999JA900477
\bibitem[Lin et al.(2002)]{2002SoPh..210....3L} Lin, R.~P., Dennis, B.~R., Hurford, G.~J., et al.\ 2002, \solphys, 210, 3. doi:10.1023/A:1022428818870
\bibitem[Liu et al.(2018)]{2018ApJ...852...10L} Liu, J., Erd{\'e}lyi, R., Wang, Y., et al.\ 2018, \apj, 852, 10. doi:10.3847/1538-4357/aa992d
\bibitem[Liu et al.(2019)]{2019FrASS...6...44L} Liu, J., Wang, Y., \& Erd{\'e}lyi, R.\ 2019, Frontiers in Astronomy and Space Sciences, 6, 44. doi:10.3389/fspas.2019.00044
\bibitem[Liu et al.(2017)]{2017ApJ...844..141L} Liu, L., Wang, Y., Liu, R., et al.\ 2017, \apj, 844, 141. doi:10.3847/1538-4357/aa7d56
\bibitem[Liu et al.(2010)]{2010ApJ...725L..84L} Liu, R., Liu, C., Wang, S., et al.\ 2010, \apjl, 725, L84. doi:10.1088/2041-8205/725/1/L84
\bibitem[Liu \& Schuck(2012)]{2012ApJ...761..105L} Liu, Y. \& Schuck, P.~W.\ 2012, \apj, 761, 105. doi:10.1088/0004-637X/761/2/105
\bibitem[Louis et al.(2014)]{2014A&A...562A.110L} Louis, R.~E., Puschmann, K.~G., Kliem, B., et al.\ 2014, \aap, 562, A110. doi:10.1051/0004-6361/201321106
\bibitem[Lu et al.(2019)]{2019ApJ...887..154L} Lu, L., Feng, L., Li, Y., et al.\ 2019, \apj, 887, 154. doi:10.3847/1538-4357/ab530c
\bibitem[Manchester et al.(2004)]{2004ApJ...610..588M} Manchester, W., Gombosi, T., DeZeeuw, D., et al.\ 2004, \apj, 610, 588. doi:10.1086/421516
\bibitem[Masson et al.(2009)]{2009ApJ...700..559M} Masson, S., Pariat, E., Aulanier, G., et al.\ 2009, \apj, 700, 559. doi:10.1088/0004-637X/700/1/559
\bibitem[Miao et al.(2019)]{2019ApJ...877...61M} Miao, Y., Liu, Y., Shen, Y.~D., et al.\ 2019, \apj, 877, 61. doi:10.3847/1538-4357/ab1a42
\bibitem[Moon et al.(2003)]{2003ApJ...588.1176M} Moon, Y.-J., Choe, G.~S., Wang, H., et al.\ 2003, \apj, 588, 1176. doi:10.1086/374270
\bibitem[Moore \& Sterling(2007)]{2007ApJ...661..543M} Moore, R.~L. \& Sterling, A.~C.\ 2007, \apj, 661, 543. doi:10.1086/516620
\bibitem[Moore et al.(2010)]{2010ApJ...720..757M} Moore, R.~L., Cirtain, J.~W., Sterling, A.~C., et al.\ 2010, \apj, 720, 757. doi:10.1088/0004-637X/720/1/757
\bibitem[Moore et al.(2013)]{2013ApJ...769..134M} Moore, R.~L., Sterling, A.~C., Falconer, D.~A., et al.\ 2013, \apj, 769, 134. doi:10.1088/0004-637X/769/2/134
\bibitem[Moore et al.(2015)]{2015ApJ...806...11M} Moore, R.~L., Sterling, A.~C., \& Falconer, D.~A.\ 2015, \apj, 806, 11. doi:10.1088/0004-637X/806/1/11
\bibitem[Moore et al.(2001)]{2001ApJ...552..833M} Moore, R.~L., Sterling, A.~C., Hudson, H.~S., et al.\ 2001, \apj, 552, 833. doi:10.1086/320559
\bibitem[Nistic{\`o} et al.(2009)]{2009SoPh..259...87N} Nistic{\`o}, G., Bothmer, V., Patsourakos, S., et al.\ 2009, \solphys, 259, 87. doi:10.1007/s11207-009-9424-8
\bibitem[Okamoto et al.(2008)]{2008ApJ...673L.215O} Okamoto, T.~J., Tsuneta, S., Lites, B.~W., et al.\ 2008, \apjl, 673, L215. doi:10.1086/528792
\bibitem[Olmedo \& Zhang(2010)]{2010ApJ...718..433O} Olmedo, O. \& Zhang, J.\ 2010, \apj, 718, 433. doi:10.1088/0004-637X/718/1/433
\bibitem[Panesar et al.(2015)]{2015ApJ...811....5P} Panesar, N.~K., Sterling, A.~C., Innes, D.~E., et al.\ 2015, \apj, 811, 5. doi:10.1088/0004-637X/811/1/5
\bibitem[Panesar et al.(2016)]{2016ApJ...822L..23P} Panesar, N.~K., Sterling, A.~C., \& Moore, R.~L.\ 2016, \apjl, 822, L23. doi:10.3847/2041-8205/822/2/L23
\bibitem[Paraschiv et al.(2020)]{2020ApJ...891..149P} Paraschiv, A.~R., Donea, A., \& Leka, K.~D.\ 2020, \apj, 891, 149. doi:10.3847/1538-4357/ab7246
\bibitem[Patsourakos et al.(2020)]{2020SSRv..216..131P} Patsourakos, S., Vourlidas, A., T{\"o}r{\"o}k, T., et al.\ 2020, \ssr, 216, 131. doi:10.1007/s11214-020-00757-9
\bibitem[Pesnell et al.(2012)]{2012SoPh..275....3P} Pesnell, W.~D., Thompson, B.~J., \& Chamberlin, P.~C.\ 2012, \solphys, 275, 3. doi:10.1007/s11207-011-9841-3
\bibitem[Prasad et al.(2020)]{2020ApJ...903..129P} Prasad, A., Dissauer, K., Hu, Q., et al.\ 2020, \apj, 903, 129. doi:10.3847/1538-4357/abb8d2
\bibitem[Raouafi et al.(2016)]{2016SSRv..201....1R} Raouafi, N.~E., Patsourakos, S., Pariat, E., et al.\ 2016, \ssr, 201, 1. doi:10.1007/s11214-016-0260-5
\bibitem[Savcheva et al.(2012)]{2012ApJ...759..105S} Savcheva, A.~S., Green, L.~M., van Ballegooijen, A.~A., et al.\ 2012, \apj, 759, 105. doi:10.1088/0004-637X/759/2/105
\bibitem[Samanta et al.(2019)]{2019Sci...366..890S} Samanta, T., Tian, H., Yurchyshyn, V., et al.\ 2019, Science, 366, 890. doi:10.1126/science.aaw2796
\bibitem[Scherrer et al.(2012)]{2012SoPh..275..207S} Scherrer, P.~H., Schou, J., Bush, R.~I., et al.\ 2012, \solphys, 275, 207. doi:10.1007/s11207-011-9834-2
\bibitem[Schmieder et al.(2013)]{2013AdSpR..51.1967S} Schmieder, B., D{\'e}moulin, P., \& Aulanier, G.\ 2013, Advances in Space Research, 51, 1967. doi:10.1016/j.asr.2012.12.026
\bibitem[Schrijver \& Title(2011)]{2011JGRA..116.4108S} Schrijver, C.~J. \& Title, A.~M.\ 2011, Journal of Geophysical Research (Space Physics), 116, A04108. doi:10.1029/2010JA016224
\bibitem[Schrijver et al.(2013)]{2013ApJ...773...93S} Schrijver, C.~J., Title, A.~M., Yeates, A.~R., et al.\ 2013, \apj, 773, 93. doi:10.1088/0004-637X/773/2/93
\bibitem[Shen et al.(2012)]{2012ApJ...750...12S} Shen, Y., Liu, Y., \& Su, J.\ 2012, \apj, 750, 12. doi:10.1088/0004-637X/750/1/12
\bibitem[Shen et al.(2011)]{2011ApJ...735L..43S} Shen, Y., Liu, Y., Su, J., et al.\ 2011, \apjl, 735, L43. doi:10.1088/2041-8205/735/2/L43
\bibitem[Shen et al.(2012)]{2012ApJ...745..164S} Shen, Y., Liu, Y., Su, J., et al.\ 2012, \apj, 745, 164. doi:10.1088/0004-637X/745/2/164
\bibitem[Su \& van Ballegooijen(2013)]{2013ApJ...764...91S} Su, Y. \& van Ballegooijen, A.\ 2013, \apj, 764, 91. doi:10.1088/0004-637X/764/1/91
\bibitem[Sittler \& Guhathakurta(1999)]{1999ApJ...523..812S} Sittler, E.~C. \& Guhathakurta, M.\ 1999, \apj, 523, 812. doi:10.1086/307742
\bibitem[Sterling et al.(2007)]{2007ApJ...669.1359S} Sterling, A.~C., Harra, L.~K., \& Moore, R.~L.\ 2007, \apj, 669, 1359. doi:10.1086/520829
\bibitem[Sterling et al.(2018)]{2018ApJ...864...68S} Sterling, A.~C., Moore, R.~L., \& Panesar, N.~K.\ 2018, \apj, 864, 68. doi:10.3847/1538-4357/aad550
\bibitem[Sterling et al.(2016)]{2016ApJ...821..100S} Sterling, A.~C., Moore, R.~L., Falconer, D.~A., et al.\ 2016, \apj, 821, 100. doi:10.3847/0004-637X/821/2/100
\bibitem[Sterling et al.(2015)]{2015Natur.523..437S} Sterling, A.~C., Moore, R.~L., Falconer, D.~A., et al.\ 2015, \nat, 523, 437. doi:10.1038/nature14556
\bibitem[Sun et al.(2012)]{2012ApJ...748...77S} Sun, X., Hoeksema, J.~T., Liu, Y., et al.\ 2012, \apj, 748, 77. doi:10.1088/0004-637X/748/2/77
\bibitem[T{\"o}r{\"o}k et al.(2004)]{2004A&A...413L..27T} T{\"o}r{\"o}k, T., Kliem, B., \& Titov, V.~S.\ 2004, \aap, 413, L27. doi:10.1051/0004-6361:20031691
\bibitem[T{\"o}r{\"o}k et al.(2011)]{2011ApJ...739L..63T} T{\"o}r{\"o}k, T., Panasenco, O., Titov, V.~S., et al.\ 2011, \apjl, 739, L63. doi:10.1088/2041-8205/739/2/L63
\bibitem[Tian et al.(2014)]{2014Sci...346A.315T} Tian, H., DeLuca, E.~E., Cranmer, S.~R., et al.\ 2014, Science, 346, 1255711. doi:10.1126/science.1255711
\bibitem[van Driel-Gesztelyi et al.(2008)]{2008AnGeo..26.3077V} van Driel-Gesztelyi, L., Attrill, G.~D.~R., D{\'e}moulin, P., et al.\ 2008, Annales Geophysicae, 26, 3077. doi:10.5194/angeo-26-3077-2008
\bibitem[Wang et al.(2017)]{2017ApJ...839..128W} Wang, J., Yan, X., Qu, Z., et al.\ 2017, \apj, 839, 128. doi:10.3847/1538-4357/aa6bf3
\bibitem[Wang et al.(2015)]{2015SCPMA..58.5682W} Wang, J., Zhang, Y., He, H., et al.\ 2015, Science China Physics, Mechanics, and Astronomy, 58, 5682. doi:10.1007/s11433-015-5682-7
\bibitem[Wang et al.(2007)]{2007SoPh..244...75W} Wang, J., Zhang, Y., Zhou, G., et al.\ 2007, \solphys, 244, 75. doi:10.1007/s11207-007-9038-y
\bibitem[Wang \& Sheeley(2002)]{2002ApJ...575..542W} Wang, Y.-M. \& Sheeley, N.~R.\ 2002, \apj, 575, 542. doi:10.1086/341145
\bibitem[Wang et al.(2018)]{2018ApJ...859..148W} Wang, Y., Su, Y., Shen, J., et al.\ 2018, \apj, 859, 148. doi:10.3847/1538-4357/aac0f7
\bibitem[Wen et al.(2006)]{2006SoPh..239..257W} Wen, Y., Wang, J., Maia, D.~J.~F., et al.\ 2006, \solphys, 239, 257. doi:10.1007/s11207-006-0181-7
\bibitem[Wheatland et al.(2000)]{2000ApJ...540.1150W} Wheatland, M.~S., Sturrock, P.~A., \& Roumeliotis, G.\ 2000, \apj, 540, 1150. doi:10.1086/309355
\bibitem[Wiegelmann(2004)]{2004SoPh..219...87W} Wiegelmann, T.\ 2004, \solphys, 219, 87. doi:10.1023/B:SOLA.0000021799.39465.36
\bibitem[Wuelser et al.(2004)]{2004SPIE.5171..111W} Wuelser, J.-P., Lemen, J.~R., Tarbell, T.~D., et al.\ 2004, \procspie, 5171, 111. doi:10.1117/12.506877
\bibitem[Yan et al.(2014)]{2014ApJ...797...52Y} Yan, X.~L., Xue, Z.~K., Liu, J.~H., et al.\ 2014, \apj, 797, 52. doi:10.1088/0004-637X/797/1/52
\bibitem[Yan et al.(2020)]{2020ApJ...904...15Y} Yan, X., Li, Q., Chen, G., et al.\ 2020, \apj, 904, 15. doi:10.3847/1538-4357/abba81
\bibitem[Yang et al.(2016)]{2016ApJ...816...41Y} Yang, B., Jiang, Y., Yang, J., et al.\ 2016, \apj, 816, 41. doi:10.3847/0004-637X/816/1/41
\bibitem[Yang \& Chen(2019)]{2019ApJ...874...96Y} Yang, B. \& Chen, H.\ 2019, \apj, 874, 96. doi:10.3847/1538-4357/ab0c9e
\bibitem[Yang et al.(2015)]{2015ApJ...803...68Y} Yang, J., Jiang, Y., Xu, Z., et al.\ 2015, \apj, 803, 68. doi:10.1088/0004-637X/803/2/68
\bibitem[Yang et al.(2012)]{2012ApJ...745....9Y} Yang, J., Jiang, Y., Zheng, R., et al.\ 2012, \apj, 745, 9. doi:10.1088/0004-637X/745/1/9
\bibitem[Yang et al.(2019)]{2019ApJ...887..239Y} Yang, L., Yan, X., Xue, Z., et al.\ 2019, \apj, 887, 239. doi:10.3847/1538-4357/ab55d7
\bibitem[Zhang et al.(2007)]{2007SoPh..241..329Z} Zhang, Y., Wang, J., Attrill, G.~D.~R., et al.\ 2007, \solphys, 241, 329. doi:10.1007/s11207-007-0229-3
\bibitem[Zhang \& Wang(2002)]{2002ApJ...566L.117Z} Zhang, J. \& Wang, J.\ 2002, \apjl, 566, L117. doi:10.1086/339660
\bibitem[Zhang et al.(2001)]{2001ApJ...548L..99Z} Zhang, J., Wang, J., Deng, Y., et al.\ 2001, \apjl, 548, L99. doi:10.1086/318934
\bibitem[Zhang \& Wang(2001)]{2001ApJ...554..474Z} Zhang, J. \& Wang, J.\ 2001, \apj, 554, 474. doi:10.1086/321343
\bibitem[Zhang et al.(2016)]{2016ApJ...827...27Z} Zhang, Q.~M., Li, D., Ning, Z.~J., et al.\ 2016, \apj, 827, 27. doi:10.3847/0004-637X/827/1/27
\bibitem[Zhou et al.(2019)]{2019ApJ...873...23Z} Zhou, G.~P., Tan, C.~M., Su, Y.~N., et al.\ 2019, \apj, 873, 23. doi:10.3847/1538-4357/ab01cf
\bibitem[Zhou et al.(2006)]{2006A&A...445.1133Z} Zhou, G.~P., Wang, J.~X., \& Zhang, J.\ 2006, \aap, 445, 1133. doi:10.1051/0004-6361:20053536
\bibitem[Zhou et al.(2020)]{2020ApJ...905..150Z} Zhou, G., Gao, G., Wang, J., et al.\ 2020, \apj, 905, 150. doi:10.3847/1538-4357/abc5b2


\end{thebibliography}
\end{document}